\documentclass[prd,twocolumn]{revtex4}
\usepackage{graphicx}
\usepackage{epsfig,color}
\usepackage{amsmath,amssymb,txfonts}

\newcommand{\ket}[1]{|{#1}\rangle}
\newcommand{\bra}[1]{\langle{#1}|}
\newcommand{\bracket}[2]{\langle#1|#2\rangle}
\newcommand{\Up}{{\uparrow}}
\newcommand{\Dn}{{\downarrow}}

\begin{document}

\title{A Formulation of Lattice Gauge Theories for Quantum Simulations}

\date{\today}

\author{Erez Zohar}
\affiliation{Max-Planck-Institut f\"ur Quantenoptik, Hans-Kopfermann-Stra\ss e 1, 85748 Garching, Germany.}
\author{Michele Burrello}
\affiliation{Max-Planck-Institut f\"ur Quantenoptik, Hans-Kopfermann-Stra\ss e 1, 85748 Garching, Germany.}

\begin{abstract}
We examine the Kogut-Susskind formulation of lattice gauge theories under the light of fermionic and bosonic degrees of freedom that provide a description useful to the development of quantum simulators of gauge invariant models. We consider both discrete and continuous gauge groups and adopt a realistic multi-component Fock space for the definition of matter degrees of freedom. In particular, we express the Hamiltonian of the gauge theory and the Gauss law in terms of Fock operators. The gauge fields are described in two different bases, based on either group elements or group representations. This formulation allows for a natural scheme to achieve a consistent truncation of the Hilbert space for continuous groups, and provides helpful tools to study the connections of gauge theories with topological quantum double and string-net models for discrete groups. Several examples, including the case of the discrete $D_3$ gauge group, are presented.

\end{abstract}

\maketitle

\section{Introduction}

The Kogut-Susskind Hamiltonian formulation \cite{KogutSusskind} of lattice gauge theories \cite{Wilson} is experiencing a renewed interest driven both by  major developments in numerical techniques, in particular based on the study of tensor networks \cite{Verstraete2004,Schuch2010}, and by groundbreaking experimental and theoretical achievements in quantum simulation, especially in the field of cold atomic gases trapped in optical lattices (see, for example, \cite{lewensteinbook}).

On the quantum simulation side, many experimental successes have been achieved recently, including the realization of artificial static gauge potentials  for both atoms trapped in harmonic potentials \cite{lin09} and in optical lattices \cite{goldman13,Aidelsburger2013,Miyake2013}, as well as the observation of Higgs modes in two-dimensional systems \cite{Endres2012}.

 Furthermore, the fast developments in the control of the interactions among atoms in an optical lattice (as well as other systems, such as trapped ions and superconducting circuits) envision the possibility of obtaining, in the near future, quantum simulations of both Abelian and non-Abelian lattice gauge theories \cite{Zohar2011,Zohar2012,Banerjee2012,Zohar2013,NA,Rishon2012,Tagliacozzo2013,AngMom,Tagliacozzo2013a,SQC,Dissipation,ZollerIons}.
 This is of particular interest, for example, for solving problems involving fermions with finite chemical potential (for example, as expected in exotic phases of QCD, such as quark-gluon plasma and color superconductivity \cite{McLerran1986,Fukushima2011}). The Euclidean lattice Monte-Carlo simulations of these
encounter the sign-problem \cite{Troyer2005} which is avoided in the framework of quantum simulations,
by the replacement of Grassman variables by real fermions. Moreover, quantum
simulations allow for real-time dynamics observation, as the Hamiltonian theory takes place in Minkowski spacetime,
unlike the statistical correlations obtained in the Euclidean approach.

Concerning the numerical and analytical study of lattice gauge theories, tensor networks allowed the investigation of the spectral properties of the $1+1$ dimensional Schwinger model with precisions comparable with the best results available from other techniques \cite{Banuls2013,Banuls2013b,Verstraete2013,Rico2014,Banuls2014} and, more in general, provide new tools to examine gauge invariant states and their dynamics  in higher dimensions as well \cite{Silvi2014,Tagliacozzo2014,Haegeman2014}.

To the purpose of obtaining a realistic lattice gauge model that can be experimentally implemented with cold atoms, it is useful to embed the Kogut-Susskind Hamiltonian in a lattice model for multi-component fermions and bosons
that realizes, as simply as possible, the required local gauge symmetry under a gauge group $G$ whose phenomenology is at quest. To accomplish this task, one of the main requirements is the possibility of building the model starting from a limited number of local degrees of freedom. In particular, we focus on characterizing the matter degrees of freedom in terms of fermionic operators which define a Fock space on each lattice vertex and describe multi-component fermions like the ones customarily used in cold atom experiments, for example in the context of lanthanide atoms like ytterbium \cite{Fallani2014} or erbium \cite{Ferlaino2014} presenting several nuclear hyperfine states which can be addressed separately. This method has already been used in several proposals for quantum simulations of lattice gauge
theories \cite{NA,AngMom,Dissipation}, applying approaches such as the prepotential formalism \cite{MathurSU2,Mathur2007} and the link model \cite{Horn1981,Orland1990,Wiese1997,Brower2004}, in which the gauge degrees of freedom are composed out of bosons or fermions, respectively. In both these approaches, the link is divided into ``left'' and ``right'' parts, with two families of such fundamental ingredients. Our approach, on the other hand, suggests considering the link as a whole piece, allowing for some mathematical simplification. It also suggests a consistent way of truncating the gauge invariant Hilbert space in a gauge invariant manner. Other differences from these two approaches are also discussed in the Appendix, exemplifying the use of the proposed method for an $SU(2)$ lattice gauge theory.

The purpose of this paper is therefore to review and reformulate the formalism of the Kogut-Susskind theory in terms of realistic fermionic and bosonic degrees of freedom. Our aim is to define the key ingredients for lattice gauge theories, with either continuous or discrete groups, in the atomic and many-body physics perspective: what must a model of interacting atoms on a lattice fulfill to properly manifest local gauge invariance?

The Fock space we adopt to describe the matter fields provides a simple platform to describe matter coupled to either finite (discrete) or continuous gauge groups, including truncated ones. In particular, we require that the internal degrees of freedom of the fermions provide a sufficient number of states to realize the smallest faithful representation of the considered group already at the single-particle level: the single-particle Hilbert space associated to a matter site must have dimension equal to that of the desired representation of the gauge group, usually the fundamental one.

With respect to the gauge field, instead, we will consider a Hilbert space whose dimension equals the group order in the case of finite groups; whereas for continuous groups we will define an efficient truncation based on its irreducible representations
(Theoretically, one could include all the possible representations by using an infinite number of
bosonic modes. This should not impose a theoretical difficulty, as quantum field theories are usually described in terms
of an infinite number of modes).

The use of a Fock space for the matter suggests a different implementation of the Kogut-Susskind Hamiltonian with respect to previous proposals aimed at simulating pure lattice gauge theories with discrete groups in Josephson junction arrays \cite{Ioffe2002,Ioffe2003,Vidal2004} or in the context of Majorana zero energy modes \cite{XuFu,Hassler2012} and parafermions \cite{Burrello2013} in topological superconductors. In these superconducting setups the charge excitations (meant as violations of the gauge constraint for the pure gauge theory) present indeed a cyclic structure in their local Hilbert space, due to the presence of a Cooper pair condensate. And the same feature appears in the study of matter-coupled gauge theories based on Majorana modes \cite{Cobanera2014}. This structure is well suited to deal with cyclic groups but it is unfeasible for cold atom implementations.

Concerning the gauge bosons, instead, the weak coupling regime of the Kogut - Susskind Hamiltonian for finite groups in the weak interaction limit can be interpreted as a topologically ordered phase. The gauge field degrees of freedom can be represented in two different bases corresponding to the elements of the gauge group and its representations. These bases allow us, indeed, to describe the weak-coupling limit of the gauge theory (for finite groups) in terms of quantum doubles \cite{Kitaev2003,Bombin2008} and string-net \cite{Levin2005,Aguado2009} models respectively.

The paper is organized as follows:  in section \ref{sec:rigid}, we shall present two ways to represent ``group states'' - either
the ``Group element basis'' or the ``Representation basis'', which provide the background for the gauge field Hilbert space. In section \ref{sec:global} we will introduce a lattice theory containing only matter (fermions) with a global gauge invariance. This gauge invariance will be lifted to a local symmetry in section \ref{sec:field}, where the full construction of the gauge field and its Hamiltonian are presented. Finally, in the Appendix,  we provide two explicit examples of this formulation of lattice gauge theories based on the finite gauge group $D_3$, and the compact Lie gauge group $SU(2)$ (where we also review Kogut and Susskind's rigid rotator derivation to provide an intuitive understanding of the gauge field Hilbert space).

Throughout this work, the Einstein summation convention (summation on double indices) is used,
for all indices but the representation ones. The representation indices may sometimes appear repeatedly, and a summation should not
be assumed for them, unless explicitly written.

\section{Representation and group element bases} \label{sec:rigid}

Let $G$ be any finite or compact Lie group. We denote its elements by $g \in G$. These may be represented by matrix representations labeled by $j$,
denoted as $D^j \left(g\right)$. We restrict ourselves to unitary representations, i.e. such that satisfy:
\begin{equation}
D^j_{mn} \left(g^{-1}\right) = D^{j*}_{nm} \left(g\right).
\end{equation}

Next, we define the ``group states''. These are labeled by the representation $j$ and an identifier within the representation, $m$ - both may be sets of indices.
These states transform (``undergo rotations'') according to their representation. If we denote the unitary operator corresponding to a group element by $\Theta_g$, then
\begin{equation} \label{transformation}
\Theta_g \left| jm \right\rangle = D^j_{nm} \left(g\right) \left| jn \right\rangle
\end{equation}

The group states are suitable for the description of the internal degrees of freedom of the matter constituents in a gauge theory: as shall be discussed in the following section, the fermionic matter carries a given representation, and may be described by a set of mutually commuting operators within it. Furthermore, by moving a fermion (matter particle) along a path on the lattice, it undergoes the transformations \eqref{transformation}, belonging to $G$, that are dictated by the local states $\ket{g}$ that the gauge fields assume in the edges along the path.

Using the above representation matrices $D^j$, which are nothing but a generalization of $SU(2)$'s Wigner matrices,
one may define a generalized, non-Abelian Fourier transform, whose coefficients (up to normalization constants) are given by:
\begin{equation}
\left\langle g | j m n \right\rangle = \sqrt{\frac{\text{dim}\left(j\right)}{\left|G\right|}} D^{j} _{mn} \left(g\right)
\label{Fourier}
\end{equation}
(For an intuitive explanation, originating from the well-known $SU(2)$ case, please refer to the Appendix).
It is thus possible to adopt two different bases for the description of the local gauge fields: a basis defined by the \emph{Group Element States} $\ket{g}$ and a basis determined by the \emph{Representation States} $\left| j m n \right\rangle$. The $D$ matrices constitute the mapping between these two bases, where the normalization follows from the great orthogonality theorem, $\text{dim}\left(j\right)$ is the dimension of the $j$th representation and $\left|G\right|$ is the order of $G$.
One may check consistency for finite groups by counting the dimensions of both spaces. The number of representation states $\left| j m n \right\rangle$ is
$\sum_j \text{dim}\left(j\right)^2$. The number of group elements (or, group element states) is the order of the group, $\left|G\right|$. According to a theorem
in the theory of finite groups, these two are equal \cite{Serre1977}.

Finally, let us consider the action of group elements on the states. Consider $U^{j}$, an \emph{operator} description of a general group element in the $j$ representation
(like a rotation matrix in $SU(2)$). As the group $G$ may, in general, be non-Abelian, one has to consider separately left and right group actions, defined using the transformation operators $\Theta^L_g$ and $\Theta^R_g$ as
\begin{equation}
\begin{aligned}
\Theta^L_g U_{mn}^{j}\Theta^{L \dagger}_g &= D_{mk}^{j}\left(g^{-1}\right)U_{kn}^{j} \\
\Theta^R_g U_{mn}^{j}\Theta^{R \dagger}_g &= U_{mk}^{j}D_{kn}^{j}\left(g\right)
\end{aligned}
\end{equation}
If $G$ is a compact Lie group, one may expand the transformation operators in terms of parameters $\alpha_g$ and generators $\mathbf{L},\mathbf{R}$,
$\Theta^L_g = e^{i \mathbf{\alpha}_g \cdot \mathbf{L}}$ and $\Theta^R_g = e^{i \mathbf{\alpha}_g \cdot \mathbf{R}}$, such that the group's algebra
 \begin{equation}
 \begin{aligned}
 \left[L_a,L_b\right] &= i f_{abc} L_c \\
 \left[R_a,R_b\right] &= i f_{abc} R_c \\
 \left[L_a,R_b\right] &= 0
 \end{aligned}
 \label{algebra}
 \end{equation}
is satisfied (where $f_{abc}$ are the group's structure constants), as well as the commutation relations
 \begin{equation}
 \begin{aligned}
 \left[L_i,U_{mn}^{j}\right]&=-\left(T^{j}_{i}\right)_{mk}U_{kn}^{j} \\
 \left[R_i,U_{mn}^{j}\right]&=U_{mk}^{j}\left(T^{j}_{i}\right)_{kn} \\
 \end{aligned}
 \label{generators}
 \end{equation}
 where $\mathbf{T}^j$ is the vector of the $j^{\rm th}$ representation matrices of the generators. The algebra and the commutation relations with the group elements may be proven from one another using the Jacobi identity.

\section{Fermions and global gauge theories} \label{sec:global}

\begin{figure}
  \centering
  \includegraphics[width=8cm]{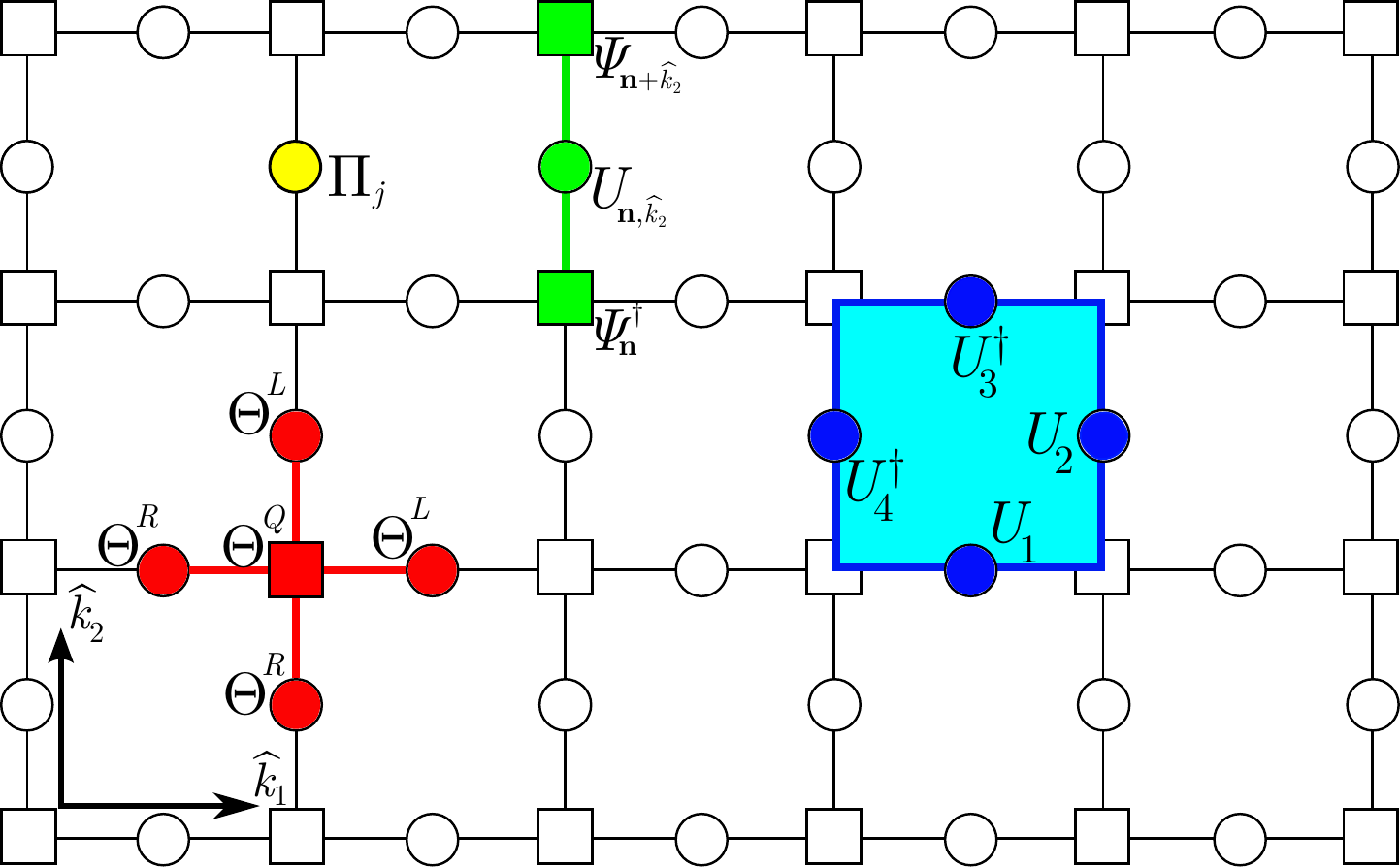}
	\caption{Graphical representation of the operators involved in the theory: the red star represents the five-site operator generating a gauge transformation \eqref{gaugetr}, the blue plaquette represents the magnetic plaquette term \eqref{mag}, the green link reproduces a vertical tunneling term \eqref{tunneling} and the yellow single-site operator is an electric term \eqref{electricp}. The squares on the vertices of the lattice correspond to matter sites, whereas the circles on the links depict gauge fields.}
  \label{fig1}
\end{figure}

\subsection{Symmetries and the Hamiltonian}
We begin, as usual, with a theory of fermions on a square lattice, having a global gauge symmetry with respect to the group $G$.
On each vertex $\mathbf{n}$ of the lattice (see Figure \ref{fig1}), we define a spinor $\psi_{\mathbf{n}}$. The spinor belongs to a given faithful representation (say, the $j$th) of $G$, and thus has $\text{dim}\left(j\right)$ components $\psi_{\mathbf{n},i}$, which we shall call group, gauge or color components. Terms like $\psi^{\dagger}_{\mathbf{n}}\psi_{\mathbf{m}}$, in which the group indices are not explicitly written,
should be understood as scalar products in group space: $\psi^{\dagger}_{\mathbf{n}}\psi_{\mathbf{m}} = \underset{i}{\sum}\psi^{\dagger}_{\mathbf{n},i}\psi_{\mathbf{m},i}$. On the other hand, the physical index $\mathbf{n}$ may be neglected when on-site properties are addressed,
as we shall do first.

A \emph{gauge transformation} on a fermionic operator, with respect to the group element $g$, is the result of acting with the operator $\Theta^{Q,j}_g$, and is defined as
\begin{equation}
\begin{aligned}
\Theta^{Q,j}_g \psi_a^{\dagger}  \Theta^{Q,j \dagger}_g &= \psi^{\dagger}_b D^{j}_{ba}\left(g\right) \,,\\
\Theta^{Q,j}_g \psi_a  \Theta^{Q,j \dagger}_g &= D^{j}_{ab}\left(g^{-1}\right)\psi_b\,.
\end{aligned}
\end{equation}

Hence we can construct a general globally gauge invariant Hamiltonian with nearest-neighbor interactions,
\begin{equation}
H = \underset{\mathbf{n}}{\sum}M_{\mathbf{n}}\psi^{\dagger}_{\mathbf{n}}\psi_{\mathbf{n}} + \underset{\mathbf{n},k}{\sum}\left(\epsilon_{\mathbf{n},k}\psi^{\dagger}_{\mathbf{n}}\psi_{\mathbf{n+\hat k}} + H.c. \right)
\end{equation}
where $k$ runs over the dimensions ($ \mathbf{\hat k}$ are the lattice vectors). One may, of course, generalize, and include many representations (types of spinors),
as long as all the group indices are contracted properly.

\subsection{The transformation operators}

\emph{Proposition 1}. The transformation operators can be written explicitly in terms of the fermionic operators. Using
\begin{equation}
q^{j}(g) = -i \text{log}\left(D^{j}\left(g\right)\right)
\end{equation}
one may define
\begin{equation} \label{theta}
\Theta^{Q,j}_g = e^{i \psi^{\dagger}_a q_{ab}(g) \psi_b} \det \left(g^{-1}\right)^{N}
\end{equation}
where $N=0$ for a vertex in the even sublattice and $N=1$ for the odd one, and
$\det \left(g^{-1}\right) \equiv \det \left(D^{j}\left(g^{-1}\right)\right)$ (with $j$ being the fermionic
matter's representation).

The part with the determinant is related to the staggering of fermions, and will be shortly explained. But first, let us prove that this, indeed, generates the required transformation.

\emph{Proof.} First, note that as we consider unitary representations, $\det \left(g^{-1}\right)^{*} = \det \left(g^{-1}\right)^{-1}$, and thus the determinant part will not contribute to the calculation of $\Theta^{Q,j}_g \psi^{\dagger}_a  \Theta^{Q,j \dagger}_g$. Thus, we define $\Theta^{Q,j}_g \equiv \tilde \Theta^{Q,j}_g \det \left(g^{-1}\right)^{N}$ and proceed without
the determinant.

Second, we wish to prove that $\tilde \Theta^{Q,j}_g$ is, indeed, a unitary operator. For that, it is sufficient to check that the exponent in Eq. \eqref{theta}, is anti-Hermitian. Note that both
$\tilde \Theta^{Q,j}_g$ and its exponent are number conserving. Thus, it is sufficient to check matrix elements of the local fermionic Fock space.
Since the representation matrices are unitary, their logarithm is anti-Hermitian and $q_{ab}=q^{*}_{ba}$.
Denote $A_{ab}=i q_{ab}$. This is anti-Hermitian: $A_{ab}=-A^{*}_{ba}$. Now we turn to the calculation of matrix elements.

Let us denote by $\left|I,n\right\rangle = \psi_{i_1}^{\dagger} ... \psi_{i_n}^{\dagger} \left|\Omega\right\rangle$ an element of the basis of the Fock space for the fermions in a given lattice site, where $\left|\Omega\right\rangle$ is the local vacuum state,
$n \leq \text{dim}\left(j\right)$ and we assume that all the participating indices are different. Then,
\begin{equation}
\begin{aligned}
& \left\langle I,n \right| \psi_a^{\dagger} A_{ab} \psi_b \left|J,n\right\rangle = \\
&= -A_{ab} \left\langle \Omega \right| \psi_{i_n} ... \psi_{i_1} \psi_b \psi_a^{\dagger} \psi_{j_1}^{\dagger} ... \psi_{j_n}^{\dagger} \left|\Omega\right\rangle + \\
&+ A_{aa} \left\langle \Omega \right| \psi_{i_n} ... \psi_{i_1} \psi_{j_1}^{\dagger} ... \psi_{j_n}^{\dagger} \left|\Omega\right\rangle = \\
&= -A_{ab} \delta^{a j_1 ... j_n}_{b i_1 ... i_n} + A_{aa}\delta^{j_1 ... j_n}_{i_1 ... i_n}
\end{aligned}
\end{equation}
where $\delta^{a j_1 ... j_n}_{b i_1 ... i_n}$ is the generalized Kronecker delta, equal to 0 if there are repeating indices in the lower/upper sets or if the two sets are different, and to $\pm 1$ if the upper sequence is respectively an even or odd permutation of the lower one.
On the other hand, and similarly,
\begin{equation}
\begin{aligned}
&\left\langle J,n \right| \psi_a^{\dagger} A_{ab} \psi_b \left|I,n\right\rangle = \\ &= -A_{ab} \left\langle \Omega \right| \psi_{j_n} ... \psi_{j_1} \psi_b \psi_a^{\dagger} \psi_{i_1}^{\dagger} ... \psi_{i_n}^{\dagger} \left|\Omega\right\rangle + \\
&+ A_{aa} \left\langle \Omega \right| \psi_{j_n} ... \psi_{j_1} \psi_{i_1}^{\dagger} ... \psi_{i_n}^{\dagger} \left|\Omega\right\rangle =\\
&= -A_{ab} \delta^{a i_1 ... i_n}_{b j_1 ... j_n} + A_{aa}\delta^{i_1 ... i_n}_{j_1 ... j_n}
\end{aligned}
\end{equation}
But now, using the fact that $A$ is anti-Hermitian, we get
\begin{equation}
\begin{aligned}
&\left\langle J,n \right| \psi_a^{\dagger} A_{ab} \psi_b \left|I,n\right\rangle = -A_{ab} \delta^{a i_1 ... i_n}_{b j_1 ... j_n} + A_{aa}\delta^{i_1 ... i_n}_{j_1 ... j_n} = \\
&= -A_{ba} \delta^{b i_1 ... i_n}_{a j_1 ... j_n} +  A_{aa}\delta^{i_1 ... i_n}_{j_1 ... j_n} =
 A_{ab}^{*} \delta^{a j_1 ... j_n}_{b i_1 ... i_n} - A_{aa}^{*}\delta^{j_1 ... j_n}_{i_1 ... i_n}= \\&= -\left\langle I,n \right| \psi_a^{\dagger} A_{ab} \psi_b \left|J,n\right\rangle^{*}
\end{aligned}
\end{equation}
and thus the exponent is anti-Hermitian and $\tilde \Theta^{Q,j}_g$ is unitary. Besides the unitarity requirement, it is also helpful from a technical aspect, since
now the inverse operator is easy to calculate using the Campbell-Baker-Hausdorff formula, resulting in
\begin{equation}
\tilde \Theta^{Q,j}_g \psi^{\dagger}_a \tilde \Theta^{Q,j \dagger}_g = \psi^{\dagger}_b \left(e^{i q^j_g}\right)_{ba} = \psi^{\dagger}_b D^{j}_{ba}\left(g\right)
\end{equation}
as required, and this completes the proof. $\square$

\subsection{Staggered fermions}

Now let us consider the staggering. Staggered fermions have been suggested by Kogut and Susskind \cite{KogutSusskind,Susskind1977} as a method to solve the problem of fermionic doubling in the continuum limit of
lattice theories. It involves the decomposition of continuum spinors into several lattice sites, such that, for example, in the case of two spin-component spinors (disregarding the gauge
components, which are not affected by staggering) the even sites correspond to particles and the odd ones to anti-particles. This allows to define a lattice analogy of the Dirac sea,
which is a state in which all the even (particle) vertices are empty, while the odd (anti-particle) vertices are full.

A meaningful staggered fermions prescription must fulfill three requirements:
\begin{enumerate}
  \item The masses have to change alternately - i.e., $M_{\mathbf{n}} = \left(-1\right)^{N_{\mathbf{n}}}M$. This contributes to the ``Dirac sea'' picture in terms of the masses.
  \item The charges may change alternately as well, depending on the gauge group.
  \item The tunneling coefficients $\epsilon$ are dictated by the continuum limit requirements, in order to obtain Dirac equation. This, of course, applies only for theories for which
  a continuum limit is expected - i.e., when $G$ is continuous.
\end{enumerate}

Let us see how the second requirement is fulfilled by the definition of the $\Theta^{Q,j}_g$ operators. Consider an empty vertex, $\left|\Omega\right\rangle$. This is not affected by
$\tilde \Theta^{Q,j}_g$, and thus its transformation law is
\begin{equation}
\left|\Omega\right\rangle \rightarrow \det \left(g^{-1}\right)^{N} \left|\Omega\right\rangle
\end{equation}
That means that an empty even vertex is invariant under the transformation (no particles - no charges), but an empty odd vertex is multiplied by $\det \left(g^{-1}\right)$
- i.e., a \emph{conjugate-trivial} transformation law.

On the other hand, what happens for a completely full site? We expect the complete opposite. Suppose $\text{dim}\left(j\right) = n$, and define
the state
\begin{equation}
\left|n\right\rangle \equiv \frac{1}{n!}\epsilon_{i_1 ... i_n}\psi_{i_1}^{\dagger} ... \psi_{i_n}^{\dagger}\left|\Omega\right\rangle
\end{equation}

\emph{Proposition 2}. Under a gauge transformation,
\begin{equation}
\left|n\right\rangle \rightarrow \det\left(g\right) \det\left(g^{-1}\right)^N \left|n\right\rangle
\end{equation}

\emph{Proof}. Let us neglect the staggering for a while, and consider only even vertices. We have:
\begin{multline} \label{gstate}
 \det\left(g\right) \left|n\right\rangle = \\ = \frac{1}{n!^2} \epsilon_{i_1 ... i_n} \epsilon_{j_1 ... j_n} D_{i_1 j_1} \cdot \cdot \cdot D_{i_n j_n}
\epsilon_{k_1 ... k_n} \psi_{k_1}^{\dagger} ... \psi_{k_n}^{\dagger}\left|\Omega\right\rangle
\end{multline}
where $\left\lbrace i\right\rbrace, \left\lbrace j\right\rbrace$ and $\left\lbrace k\right\rbrace$ are three permutations of the gauge indices and the matrix $D$ is the representation of the group element $g$ in the faithful representation of the fermions.

We use the identity $\epsilon_{i_1 ... i_n}\epsilon_{k_1 ... k_n} = \delta^{i_1 ... i_n}_{k_1 ... k_n}$ to eliminate two Levi-Civita symbols. Summing over $\left\lbrace k\right\rbrace$ there are $n!$ summands, and the generalized delta is either $1$ or $-1$ for each of them according to $\left\lbrace k\right\rbrace$ and $\left\lbrace i\right\rbrace$ being even or odd permutations one of the other. Accounting also for the anticommutation relations of the $\psi^\dag$ we obtain:
\[
 \frac{1}{n!}\delta^{i_1 ... i_n}_{k_1 ... k_n} \psi_{k_1}^{\dagger} ... \psi_{k_n}^{\dagger} = \psi_{i_1}^{\dagger} ... \psi_{i_n}^{\dagger}.
\]
Substituting this relation in the Eq. \eqref{gstate} we have:
\begin{equation}
\det\left(g\right) \ket{n} = \frac{1}{n!} \epsilon_{j_1 ... j_n}  D_{i_1 j_1} \cdot \cdot \cdot D_{i_n j_n}
 \psi_{i_1}^{\dagger} ... \psi_{i_n}^{\dagger}\left|\Omega\right\rangle
\end{equation}
which is exactly the result of a transformation on each fermionic operator separately. For odd vertices just multiply by the missing determinant. $\square$

Thus, we get that a fully occupied even vertex undergoes the transformation as an empty odd vertex and vice versa - as expected from the Dirac sea picture.

As another example of the charge conjugation due to staggering, consider a site occupied by a single fermion,
\begin{equation}
\left|  a \right\rangle \equiv \psi^{\dagger}_a \ket{\Omega}
\end{equation}

There we get
\begin{equation}
\left|  a \right\rangle \rightarrow
D^{j}_{ba}\left(g\right) \det \left(g^{-1}\right)^{N} \left|b\right\rangle
\end{equation}
This has to be compared to the state
$\left| a' \right\rangle \equiv \psi_a \left|n\right\rangle$, for which:
\begin{equation}
\begin{aligned}
& \left| a' \right\rangle  \rightarrow D^{j}_{ab}\left(g^{-1}\right) \det\left(g\right) \det\left(g^{-1}\right)^N \left|b'\right\rangle = \\
&= D^{j*}_{ba}\left(g\right) \det\left(g\right)^{1-N} \left|b'\right\rangle
\end{aligned}
\end{equation}
where the charge conjugation may be explicitly seen.

\subsection{Some examples}

As final remarks for this section, we shall consider examples for several gauge groups.

Let us consider first the smallest discrete non-Abelian group, the dihedral group $D_3=S_3$, corresponding to the symmetry group of the equilateral triangle or, equivalently, to the permutations of three elements. The group has six elements and the only possible faithful irreducible representation has dimension 2. In this case, $\det\left(g\right) = \pm 1$ and we obtain that a doubly occupied odd site,
as well as an empty even site, undergo the transformation in the trivial representation (without a determinant), whereas the opposite cases undergo it in the parity representation (with a determinant). In this case, $\left|a'\right\rangle = \epsilon_{ab}\left|b\right\rangle$, and one can easily verify that an even $\left|a\right\rangle$ undergoes the conjugate  transformation of an odd $\left|a'\right\rangle$, and vice versa.

Another important case is the one of compact Lie groups. Consider $U(N)$ or $SU(N)$ for example,
 for which infinitesimal transformations can be defined, and the group representations can be written in terms of Hermitian generators $\mathbf{T}^{j}$ in the form $D^{j}_{ab}\left(g\right) = e^{i \mathbf{\alpha}_g \cdot \mathbf{T}^{j}}$. In this way we can define:
\begin{equation} \label{chargedef}
\left(q^{j}_g\right)_{ab} =  \mathbf{\alpha}_g \cdot \mathbf{T}^{j}_{ab}
\end{equation}
and one can easily extract the Hermitian gauge charge $Q^{j}_i$, defined (as usual) by:
\begin{equation}
\Theta^{Q,j}_g = e^{i \mathbf{\alpha}_g \cdot \mathbf{Q}^{j}}.
\end{equation}

For $SU(N)$ all the determinants are 1, thus the staggering plays no role in the charge definition, and we get that the charges are
\begin{equation}
\mathbf{Q}^{j}_{SU(N)} = \psi^{\dagger}_a \mathbf{T}^{j}_{ab} \psi_b \label{NAcharge}
\end{equation}
as in \cite{KogutSusskind}. This defines a fermionic generalization of the Schwinger algebra \cite{Schwinger}.

For $U(1)$, on the other hand (or the Abelian subgroup of $U(N)$), we have:
\begin{equation}
\Theta^{Q,j}_{\phi} = e^{i \left(\psi^{\dagger} \psi - N \right)\phi}.
\end{equation}
This results in the well-known Abelian staggered charge:
\begin{equation}
Q_{U(1)} = \psi^{\dagger} \psi - \frac{1}{2}\left(1-\left(-1\right)^N\right).
\end{equation}

Thus, for $U(N)$ there are both Abelian and non-Abelian charges corresponding to the $Q_{U(1)}$ and $\mathbf{Q}^{j}_{SU(N)}$ presented above.

\section{Local gauge symmetry: the gauge fields} \label{sec:field}

\subsection{Inclusion of a gauge field}

We now have all the ingredients for lifting the gauge symmetry to be local. For that, one simply has to introduce ``connections'' - group elements of the form $U^{j}$ presented earlier on every link (see Figure \ref{fig1}), where the representation $j$ corresponds to the one chosen for the fermions. The modified Hamiltonian is thus
\begin{equation}
H = \underset{\mathbf{n}}{\sum}M_{\mathbf{n}}\psi^{\dagger}_{\mathbf{n}}\psi_{\mathbf{n}} + \underset{\mathbf{n},k}{\sum}\left(\epsilon_{\mathbf{n},k}\psi^{\dagger}_{\mathbf{n}} U^{j}_{\mathbf{n},k}\psi_{\mathbf{n+\hat k}} + H.c. \right)
\label{Hint}
\end{equation}
where summation on internal (group) indices is, again, assumed:
\begin{equation} \label{tunneling}
\psi^{\dagger}_{\mathbf{n}} U^{j}_{\mathbf{n},k}\psi_{\mathbf{n+\hat k}} = \left(\psi^{\dagger}_{\mathbf{n}}\right)_a \left(U^{j}_{\mathbf{n},k}\right)_{ab}
\left(\psi_{\mathbf{n+\hat k}} \right)_b
\end{equation}
The staggering procedure is unaffected by this modification.

The operators $U^j$ act on the links of the square lattice which are characterized by a Hilbert space identical to the one described in section \ref{sec:rigid}: on each link there is a Hilbert space as the one defined in \ref{sec:rigid}, playing the role of the gauge field in the theory. Thus it is easy to verify that this model
is, indeed, local gauge invariant - i.e., the gauge transformation involves different group elements in different vertices of the lattice. Define on each vertex the transformation
\begin{equation} \label{gaugetr}
\Theta_g = \underset{o}{\prod}\Theta^L_{g,o}\underset{i}{\prod}\Theta^R_{g,i}\Theta^Q_{g}
\end{equation}
with $i,o$ being the links ingoing and outgoing to/from the vertex. This is the generator of gauge transformations. Acting
on a $U^{j}$ operator with $g$ on the left and $h$ to the right results in
\begin{equation}
U^{j}_{mn} \rightarrow D^{j}_{mm'}\left(g^{-1}\right) U^{j}_{m'n'} D^{j}_{n'n}\left(h\right)
\end{equation}
Combining it with the fermionic part, which we already know, we see that this Hamiltonian is, indeed, gauge invariant.

Thus, since for each vertex $\mathbf{n}$ we have
$\left[H,\Theta_{g,\mathbf{n}}\right] = 0$, $\Theta_{g,\mathbf{n}}$ is a local symmetry, decomposing the Hilbert space of the theory into separate sectors, corresponding to eigenvalues of these operators. These sectors correspond to static charge configurations. Let us set them all to 1, i.e., for every $\mathbf{n}$, any physical state $\left|phys\right\rangle$ is invariant by the local transformations:
\begin{equation}
\Theta_{g,\mathbf{n}} \left|phys\right\rangle = \left|phys\right\rangle
\end{equation}
This is Gauss's law which defines the physical, gauge-invariant subspace of the Hilbert space. In case of continuous groups, one may rephrase the Gauss law in terms of generators and charges,
\begin{equation} \label{gausscharge}
\mathbf{G}_{\mathbf{n}} = \underset{i}{\sum} \mathbf{R}_{g,i} + \underset{o}{\sum}\mathbf{L}_{g,o} + \mathbf{Q}_{\mathbf{n}}
\end{equation}
where the generators $\mathbf{L}$ and $\mathbf{R}$ fulfill the algebra relations \eqref{algebra} and $\mathbf{Q}$ in defined in \eqref{chargedef}. Eq. \eqref{gausscharge} constitutes Gauss's law for all the charges in the theory and, in particular, in the case of Abelian charges, one can easily verify that $\mathbf{R} = -\mathbf{L}$, and thus a discrete version of the divergence of the generators is obtained. Therefore, we may identify the group generators with an electric field, whereas their conjugate variables constitute the vector potentials.
Having both right and left generators in non-Abelian groups thus corresponds to having right and left electric fields, due to the fact that in such theories the link -
or, in other words, the gauge field - is charged under the gauge group, unlike in Abelian theories (compare the charged $SU(3)$ gluons with the chargeless $U(1)$ photon, for example).
Generalizing this to any valid group $G$, we deduce that the representation basis is appropriate for describing charges and electric fields, while the group element basis is most suited, for example, for the discussion of magnetic vortices in topological models.

\subsection{The gauge field dynamics}

To complete the picture, we wish to make the gauge field dynamic. This is done in the usual way, by adding the simplest local and gauge-invariant terms, involving the gauge field only, to the Hamiltonian.
The first term it is possible to introduce is constituted by string operators which change the state of the gauge bosons on a closed path by adding a loop of electric flux; in particular these terms may be written in terms of Wilson loops - traces of products of group elements $U^{j}$. Locality leads us to choose
the smallest closed paths to obtain operators acting on each lattice plaquette $p$. With these operators we define the \emph{magnetic part} of the Hamiltonian,
\begin{equation}
H_B = -\frac{1}{2\varg^2} \sum_p \left( \text{Tr}\left(U^{j}_1 U^{j}_2 U^{j \dagger}_3 U^{j \dagger}_4\right) + H.c. \right)
\label{mag}
\end{equation}
where the numbers $1-4$ are taken according to the plaquette orientation convention presented in Figure \ref{fig1}. The representation $j$ is the same along all links (otherwise the matrix product in group space is not defined). In $SU(N)$ theories, for example,  one  usually chooses the fundamental representation, but other choices are possible too. For the group $D_3$, instead, there is only one faithful irreducible representation and thus the matrices (and the matter fields) are in it.

$\varg$ is the gauge field coupling constant (such as the electron charge for quantum electrodynamics). The name ``magnetic Hamiltonian'' is due to the fact that this term generates the magnetic energy in the continuum limit for compact Lie groups. This is a well-known fact, and we
shall review it explicitly in the group element representation later on.

Furthermore, the plaquette operator $W=U^{j}_1 U^{j}_2 U^{j \dagger}_3 U^{j \dagger}_4$ plays the role of an Aharonov-Bohm transformation that a fermion, moving in a closed loop around the plaquette, undergoes. This allows an interpretation of the eigenvalues of $W$ in terms of the presence of different magnetic vortices on the plaquette.
In particular the previous plaquette operator commutes with Gauss's law, therefore, it is possible to describe the physical subspace of the theory in terms of the eigenstates of all the plaquette operators $W$; namely, the eigenvalues of $\text{Tr}W$ represent all the possible magnetic charges in the theory which correspond to the conjugacy classes of the group $G$.

The magnetic terms in the Hamiltonian constitute a self interacting term for the gauge field, but it is not sufficient to define its dynamics, since the operators $W$ commute with both Gauss's law and the tunneling terms \eqref{tunneling}.

The dynamics of the field must be introduced by adding in the Hamiltonian interactions that depend on the conjugate variables of group elements - which are the
representations. By considering single-link terms only, the most general interaction of this kind may be written in the form
\begin{equation} \label{electricp}
H_E = \frac{\varg^2}{2} \underset{links}{\sum}\underset{j}{\sum}\mathcal{E}\left(j\right) \Pi_j
\end{equation}
where $\Pi_j$ is the projector onto the $j$ representation, and $\mathcal{E}\left(j\right)$ is a function of the representation only.
$H_E$ commutes with Gauss's law but does not commute with the magnetic part, thus providing a non-trivial dynamics to the gauge field also in the absence of matter.

$H_E$ can be regarded as the \emph{electric part} of the
Hamiltonian. In order to see that, consider $G=SU(N)$, for example. Then a proper choice would be $\mathcal{E}\left(j\right) = C_2 \left(j\right)$ - i.e., $H_E$ is a sum of local quadratic
Casimir operators (for $SU(2)$, e.g., $\mathcal{E}\left(j\right) = j\left(j+1\right)$). Then we get
\begin{equation}
H_E = \frac{\varg^2}{2} \underset{links}{\sum} \mathbf{E}^2
\label{HESUN}
\end{equation}
with the quadratic Casimir operator $\mathbf{E}^2 = \mathbf{L}^2 = \mathbf{R}^2$. And we recognize the sum over the squares of the electric field all over the lattice as the well known
form of electric energy.

Thus, the Hamiltonian we have for the gauge field,
\begin{equation}
H_{KS} = H_E + H_B
\end{equation}
is a generalization of the well-known Kogut-Susskind Hamiltonian \cite{KogutSusskind,Kogut1979,Kogut1983} for lattice gauge theories with compact Lie gauge groups. The Kogut-Susskind Hamiltonian, its corresponding Wilson
action \cite{Wilson} and their continuum limit are the sources of the $\varg^2$ factors in $H_E$ and $H_B$.

As a final remark, note that in the case that $G$ is Abelian, or has an Abelian subgroup ($U(1)$ or $\mathbb{Z}_N$), one can add a term corresponding to the Abelian charge in the electric
energy. For example, the square of the electric field in $U(1)$, or the Abelian electric field in $U\left(N\right)$ for $N>1$.

\subsection{Realization in representation space}

Analogously to the construction of the fermionic transformation operators, we would like to obtain the group elements $U^{j}$ and the transformation operators $\Theta^L,\Theta^R$ in terms of operators acting within the gauge field local Hilbert space.

We recall from section \ref{sec:rigid}, that this Hilbert space consists of the states $\left| j m n \right\rangle$.
 Several proposals to simulate lattice gauge theories in ultracold atomic setups rely on the simultaneous presence of different atomic species to mimic the matter and gauge fields in the theory \cite{Zohar2013,NA,Rishon2012,AngMom,Dissipation}. The atoms are trapped in optical lattices \cite{lewensteinbook}, which define potentials that are sensitive to the internal levels of the atoms, usually constituted by different hyperfine levels. In these proposals, fermionic atoms simulating the matter are trapped on the vertices of the simulated lattice, while other atomic species (either bosonic or fermionic), simulating the gauge field, are trapped on the links.
 Optical lattices with the required geometry can be experimentally obtained by exploiting the simultaneous presence of lasers with commensurate wavelengths, which allow to impose a superlattice potential, i.e. to superimpose multiple optical potentials (see, for example, \cite{Jo2012,Sengstock2013}). In this way, one can engineer superlattices involving different optical potentials experienced by different atomic species which are therefore trapped in different locations, e.g. on the links and the vertices of some spatial lattice, or on both of them.

In such schemes, the fermionic atoms simulating the matter are described in second quantization, using a Fock space, as in the simulated theories: a fermion in the gauge theory - for example, an electron or a quark - is simulated by a fermionic atom with a similar second quantization representation. The gauge bosons, however, are described in a slightly different manner. In the simulated theory, in fact, one does not need a local Fock space for them: the use of a simple Hilbert space consisting of the states $\left| j m n \right\rangle$ is sufficient. In the atomic picture, however, one may represent each of these states by a different internal level (atomic species) of a single atom occupying the link. Thus, the local space of Hilbert states $\left| j m n \right\rangle$ has to be lifted into a local $\emph{Fock space}$, determined by the modes $a^{j \dagger}_{mn}$, defined as
  \begin{equation}
a^{j \dagger}_{mn} \left|\Omega\right\rangle = \left| j m n \right\rangle.
\end{equation}
where $\left|\Omega\right\rangle$ is the local atomic vacuum (on the link). One should note that the simulated ``strong coupling'' vacuum (for example, the $SU(2)$ singlet state $\left| 0 0 0 \right\rangle$), is not the atomic vacuum $\left|\Omega\right\rangle$. In fact, according to the above equation, $\left| 0 0 0 \right\rangle = a^{0 \dagger}_{0} \left|\Omega\right\rangle$. This is due to the simple fact that we are mapping a Hilbert space into a Fock space. One could ask, then, whether this introduces a redundant representation, or, even worse, more states than required. In order to avoid that, one must make sure that the total local atomic number, $N_{tot} = \underset{jmn}{\sum} a^{j \dagger}_{mn}a{j}_{mn}$ is always one. In this way, by imposing the presence of a single atom per link, we select a subspace of the Fock space which behaves like a simple Hilbert space.

This requirement is fulfilled when the Hamiltonian commutes with the local number operator $N_{tot}$ and the initial state contains one atom on each link. That holds as long as the Hamiltonian is constructed out of local number conserving terms - i.e., functions of $a^{j \dagger}_{mn}a^{j'}_{m'n'}$. This is also required by the atomic number conservation: the atomic physics is nonrelativistic, therefore the atomic interactions are only number conserving
\cite{footnote2}. Thus, we shall construct here all the operators relevant for expressing the gauge field degrees of freedom ($U^{j},\Theta^L,\Theta^R,\Pi_j$) using these local Fock space operators, considering only local number conserving terms.

As a final remark, before starting the construction, note that this choice of the Fock-subspace implies that the atoms simulating the gauge field may be either bosonic or fermionic.
For the sake of simplicity and without any loss of generality, we will call these gauge particles ``bosons'', irrespective of their nature, in agreement with the usual bosonic representation of gauge fields.

Let us start with the construction of $U^{j}$ in terms of the operators $a,a^\dag$. Motivated by equation (\ref{Fourier}), we can deduce that in group element space,
\begin{equation} \label{gUh}
\left\langle g \right| U_{mn}^{j} \left| h \right\rangle = D^{j}_{mn}\left(g\right) \delta_{gh}\,,
\end{equation}
see \cite{Tagliacozzo2014} as well.
From this starting point, we may derive the representation space form of  $U^{j}$ (later we will show the other way around, as a consistency check).

\emph{Proposition 3}.
The $ U^{j} $ matrix elements are given by
\begin{equation}
 U_{mm'}^{j} = \underset{J,K}{\sum} \sqrt{\frac{\text{dim}\left(J\right)}{\text{dim}\left(K\right)}}  \bracket{JMjm}{KN} \left\langle KN' | JM'jm' \right\rangle
 a^{\dagger K}_{NN'} a^{J}_{MM'}
 \label{CGS}
\end{equation}
where $J,K$ are representations of $G$, the sum over the related gauge components $M,M',N$ and $N'$ is understood hereafter, and $\left\langle KN | JMjm \right\rangle,\, \left\langle KN' | JM'jm' \right\rangle$ are Clebsch-Gordan coefficients.

This operator changes, in general, the representation associated to the gauge field; in other words, it changes the electric flux carried by the link, acting like an exponential of the vector potential as expected from the definition of the tunneling term \eqref{tunneling}.

\emph{Proof.}
We wish to calculate the matrix elements of $U_{mm'}^{j}$ in the representation basis, using equations (\ref{gUh}) and (\ref{Fourier}). Using the Clebsch-Gordan series
\cite{footnote,brink}, one obtains
\begin{equation}
\begin{aligned}
& \left\langle KNN' \right| U_{mm'}^{j} \left| JMM' \right\rangle =
\frac{\sqrt{\text{dim}\left(J\right) \text{dim}\left(K\right)}}{\left|G\right|} \times \\
& \times \underset{g,I}{\sum}D^{K*}_{NN'}\left(g\right)D^{I}_{LL'}\left(g\right) \bracket{JMjm}{IL} \left\langle IL' | JM'jm' \right\rangle
\end{aligned}
\end{equation}
Using the great orthogonality theorem for summation over the group elements, one obtains
\begin{equation} \label{Uj}
\left\langle KNN' \right| U_{mm'}^{j} \left| JMM' \right\rangle = \sqrt{\frac{\text{dim}\left(J\right)}{\text{dim}\left(K\right)}}  \bracket{JMjm}{KN} \left\langle KN' | JM'jm' \right\rangle
\end{equation}
which proves the proposition. $\square$

Next, let us consider the transformation operator.

\emph{Proposition 4}. The left and right transformation operators are respectively:
\begin{align}
\Theta_g^{L} &= \underset{j}{\sum} a^{\dagger j}_{ml}  D_{mn}^{j*} \left(g\right) a^{j}_{nl}\,, \label{Ltransf}\\
\Theta_g^{R} &= \underset{j}{\sum} a^{\dagger j}_{lm}  D_{mn}^{j} \left(g\right) a^{j}_{ln}\,. \label{Rtransf}
\end{align}

\emph{Proof.}
First, let us check that these operators correctly implement the composition rules of the transformations. Since these operators act block-diagonally in the representations, it is sufficient to prove it for a generic representation $j$.
 Thus,
 \begin{equation}
 \Theta_g^{L,j} \Theta_h^{L,j} = a^{\dagger j}_{ml}  D_{mn}^{j*} \left(g\right) a^{j}_{nl} a^{\dagger j}_{ac}  D_{ab}^{j*} \left(h\right) a^{j}_{bc}
 \end{equation}
However, $a^{j}_{nl} a^{\dagger j}_{ac} = \delta_{na}\delta_{lc} \mp a^{\dagger j}_{ac}a^{j}_{nl}$, where the $\mp$ sign holds for bosons or fermions respectively. The second term vanishes due to the multiplication by $a^{j}_{bc}$
(recall that there is only a single particle on the link at a time). Thus we get a product of the representation matrices which leads to the desired result:
 \begin{equation}
 \Theta_g^{L,j} \Theta_h^{L,j} = a^{\dagger j}_{ml}  D_{mn}^{j*} \left(gh\right) a^{j}_{nl} = \Theta_{gh}^{L,j}
 \end{equation}
 Also, note that thanks to the unitarity of the representations, $\Theta_{g^{-1}}^{L} = \Theta_g^{\dagger L}$. And thus, by the former argument,
 \begin{equation}
 \Theta_g^{L} \Theta_g^{\dagger L} = \Theta_1^{L} = \underset{j} {\sum} a^{\dagger j}_{lm} a^{j}_{lm} = 1
 \end{equation}
 using the constraint of exactly one particle per link, unitarity is proven. Similar arguments lead to
 $\Theta_g^{R,j} \Theta_h^{R,j} = \Theta_{gh}^{R,j}$ and $\Theta_g^{R} \Theta_g^{\dagger R} = 1$.

Finally, one has to verify the transformation law, i.e. whether
\begin{equation}
\begin{aligned}
&\left\langle K N N' \right| \Theta_g^{R} U_{mn}^{j} \Theta_g^{\dagger R} \left| J  M  M' \right\rangle = \\
&\left\langle K N N' \right|  U_{mk}^{j} D^{j}_{kn}\left(g\right) \left| J  M  M' \right\rangle
\end{aligned}
\end{equation}
and similarly for the left transformations.
Calculating the LHS, and disregarding the normalization factor and the left Clebsch-Gordan, which are not affected by the transformation, one obtains
\begin{equation}
\begin{aligned}
& \left\langle K \tilde N' | J \tilde M' j n \right\rangle D^{K}_{N' \tilde N'} \left(g\right) D^{J}_{\tilde M'  M'} (g^{-1}) = \\
& \left\langle K N' \right| \Theta_g \left| K \tilde N' \right\rangle \left\langle K \tilde N' | J \tilde M' j n' \right\rangle
 \left\langle J \tilde M' j n' \right| \left( \Theta_g^{\dagger} \left|JM' \right\rangle \otimes  \left|jn \right\rangle \right)= \\
& \left\langle K N' \right| \left( \left|JM' \right\rangle \otimes \Theta_g \left|jn \right\rangle \right)= \\
& \left\langle K N' | J M' j k \right\rangle \left\langle J M' j k \right| \left( \left|JM' \right\rangle \otimes \Theta_g \left|jn \right\rangle \right)= \\
& \left\langle K N' | J M' j k \right\rangle D^{j}_{kn}\left(g\right)
\end{aligned}
\end{equation}
where completeness had been used.
This completes the proof for the right transformations, and similar arguments for the left ones complete everything.  $\square$

Let us now perform a consistency check and see what we get for the case of a compact Lie $G$, $U(N)$ or $SU(N)$, for example.
In this case, for example,
\begin{equation}
\Theta_g^{L} = \sum_j a^{\dagger j}_{ml}  e^{-i \mathbf{\alpha}_g \cdot \mathbf{T}^{j*}}  a^{j}_{nl} = \sum_j  e^{-i a^{\dagger j}_{ml}  \mathbf{\alpha}_g \cdot \mathbf{T}^{j*} a^{j}_{nl}}
\end{equation}
(the creation and annihilation operators may be lifted to the exponent due to the fact we have a single particle in our Hilbert space).
On the other hand, we expect the relation
\begin{equation}
\Theta_g^{L} = e^{i \mathbf{\alpha}_g \cdot \mathbf{L}}
\end{equation}
 and thus we can extract the generators
 \begin{equation}
\mathbf{L} = - \underset{j}{\sum} a^{\dagger j}_{ml} \mathbf{T}_{nm}^{j} a^{j}_{nl} \label{Lgen}
\end{equation}
(note that as the $T$ matrices are hermitian, $T^{*} = T^{T}$). Similarly, one obtains
 \begin{equation}
\mathbf{R} =  \underset{j}{\sum} a^{\dagger j}_{lm} \mathbf{T}_{mn}^{j} a^{j}_{ln} \label{Rgen}
\end{equation}
These two sets of operators satisfy the required algebra (\ref{algebra}) and commutation relations (\ref{generators}). These may be proven either directly,
or by the use of an infinitesimal transformation and the Jacobi identity.

Finally, let us show explicitly that this formulation allows us, indeed, to obtain all the physical states of the Kogut-Susskind model.
The Hamiltonian we wish to realize consists of three parts, described by equations
(\ref{Hint}), (\ref{mag}) and (\ref{electricp}).
Gauge invariance is ensured due to the constructions shown above: first, the Hamiltonian for the matter (\ref{Hint}) and the magnetic term (\ref{mag}) are constructed by exploiting the operators $U^j_{mn}$ introduced in Eq. (\ref{CGS}), which are, as shown above, equivalent to the same operators in the model of Kogut and Susskind, although written in terms of the new bosonic modes introduced in this work. The equivalence holds only in the case of a constant $N_{\text{tot}}=1$ on each link. However, the $U^j_{mn}$ operators consist of linear combinations of the operators $a^{\dagger K}_{NN'} a^{J}_{MM'}$ and commute with $N_{\text{tot}}$, thus we conclude that the gauge-matter interaction conserves $N_{tot}$. As for the electric part, Eq. (\ref{electricp}), if one makes the usual choice of Casimir operators for $SU(N)$, see Eq. (\ref{HESUN}), the Kogut-Susskind electric term is achieved, and it is straightforward to see that, since the generators $\mathbf{L}$ and $\mathbf{R}$ are number-conserving (see Eqs. (\ref{Lgen}) and (\ref{Rgen})), the desired result is obtained. Otherwise, considering the more general case, the projection operators to a particular $j$ representation, required in (\ref{electricp}), are given by
\begin{equation}
\Pi_j = \underset{mn}{\sum}a^{j \dagger}_{mn}a^{j}_{mn}\,,
\end{equation}
where the sum includes all the number operators of a single representation. Within the subspace of interest, each of the number operators has only the eigenvalues 0 or 1, and so does the operator $\Pi_j$, therefore it is a projection operator, $\Pi_j^2=\Pi_j$. These operators commute with $N_{tot}$ and thus we finally deduce that so does the full Hamiltonian; therefore the local atomic numbers are conserved by the dynamics.

Following Kogut and Susskind \cite{KogutSusskind}, all the possible physical (gauge-invariant) states are achieved using gauge-invariant singlet operators
(with the group indices completely contracted). These are either closed flux loops of $U^j_{mn}$ operators, or strings of these operators with fermionic operators on the edges.
The physical states are obtained by acting with these operators on $\left|vac\right\rangle$ , i.e. the strong coupling gauge field vacuum (the singlet state $\left|000\right\rangle$ on all the links) and the Dirac sea for the fermions \cite{KogutSusskind}. In the proposed formulation, $\left|vac\right\rangle$ consists of the appropriate fermionic configuration, and the state $a^{0\dagger}_{00}\left|\Omega\right\rangle$ on each and every link. This state is in the desired subspace: $N_{\text{tot}}=1$ everywhere. Next, all the other physical states may be constructed. The singlet loops/strings consist of operators which are realizable in our approach, and furthermore conserve the $N_{\text{tot}}=1$ of the initial state $\left|vac\right\rangle$. Therefore all the physical states included in the Kogut-Susskind model are also included in our description, as well as their dynamics, since, as argued above, one can also express the Hamiltonian using the proposed approach. Since the Hilbert space of states, the Hamiltonian and the symmetry are equivalent, one may deduce that the proposed approach is a full formulation of the Kogut-Susskind model.

An important property of this representation is the ability to truncate the $U^j_{mn}$ operators: note that each of the summands
in (\ref{CGS}) is actually gauge invariant on its own, and thus one could, for example, consider only $J \leq J_{max}$ and truncate the series in a gauge invariant manner. The result will be, indeed, a non-unitary $U^j_{mn}$ operator, but
yet it will be gauge invariant with all the required transformation properties. As a consequence, one will have to truncate accordingly (such that $J \leq J_{max}$) the transformation operators of proposition 4 as well.
One must make sure that all the chosen representations are fully included (i.e., all the $\left|jmn\right\rangle$ states of any $j\leq J_{max}$ are included), that the selected representations may be connected by the $U^j_{mn}$ operators in the representation included in the Hamiltonian, and that the trivial representation is also included and connected.
Then the possible physical states will be the components of the full Kogut-Susskind physical superpositions which involve the selected representations. These will be constructed again out of $\left|vac\right\rangle$ in a way that conserves $N_{\text{tot}}=1$. Note that these are approximate eigenstates of the the full Kogut-Susskind Hamiltonian, as the Hilbert space is a truncated one. Thus one may see it as a cutoff (in representation space) of the strong coupling expansion. For that reason it might be unsuitable for weak limit purposes (which, indeed, require the group element representation as we discuss in the next section). For example, in \cite{Orland1990} it was shown that similar models result in ``non-relativistic gluons''.

\subsection{Realization in group element space}

Better insight about the magnetic part of the Hamiltonian can be gained by exploiting the space of group elements. First, we shall convert, as a consistency check, the $U^{j}$ from the previous subsection to this basis.
Using (\ref{Fourier}) and completeness relations, one obtains that:
\begin{equation}
\left\langle g \right| U^{j}_{mn} \left| h \right\rangle = \frac{1}{\left|G\right|}\sum_J\textrm{dim}\left(J\right) \textrm{Tr}\left(D^{J}\left(g^{-1}h\right)\right)
D^{j}_{mn} \left(g\right).
\end{equation}
Note that $\sum_J\textrm{dim}\left(J\right)\text{Tr}\left(D^{J}\left(g^{-1}h\right)\right)$ is the character of the regular representation of $g^{-1}h$, which is equal to $|G|$ if $g^{-1}h=1$ and vanishes otherwise \cite{Serre1977}. Thus we recover, unsurprisingly, Eq. \eqref{gUh}.

A straightforward calculation allows us to rewrite $H_B$ in the form:
\begin{widetext}
\begin{equation} \label{HB}
H_B = -\frac{1}{2\varg^2} \sum_{p}\sum_{g_1g_2g_3g_4}\left(\textrm{Tr}\left[D^{j}\left(g_1 g_2 g^{-1}_3 g^{-1}_4 \right)\right]
\left| g_1 g_2 g^{-1}_3 g^{-1}_4 \right\rangle \left\langle g_1 g_2 g^{-1}_3 g^{-1}_4 \right| + H.c. \right).
\end{equation}
\end{widetext}
This equation shows that the plaquette term is constituted by the sum of the projectors over all the possible group elements obtained as the oriented product of the transformations $g_i$ associated to the plaquette edges. Each projector in the sum is weighted with the character $\chi_j\left(C\right)=\textrm{Tr}\left[D^{j}\left(g_p\right)\right]$ of the related group element in the $j$th (usually the fundamental) representation.  Thus, the product of the elements along the square edges, $g_1 g_2 g^{-1}_3 g^{-1}_4 \equiv g_p$, constitutes the transformation that a matter particle undergoes when it moves around the plaquette. For non-Abelian groups, $g_p$ is mapped into $h^{-1}g_p h$ by a gauge transformation $\Theta_h$ applied on the first vertex of the plaquette. Therefore, to rewrite the projectors appearing in Eq. \eqref{HB} in a gauge-invariant form, we must consider the effect of these gauge transformations; in particular, by exploiting the gauge invariance of $\textrm{Tr}\left[D^{j}\left(g_p\right)\right]$, the plaquette operator may be expressed as:
\begin{equation}
H_B = -\frac{1}{2 \varg^2} {\sum}_{p}\underset{C}{\sum}\chi_j\left(C\right) \Pi_{C,p} + H.c.
\label{discmag}
\end{equation}
where $C \equiv \left\lbrace h^{-1}g_C h\,, h\in G  \right\rbrace$  labels the conjugacy classes of the group $G$, and $\Pi_{C,p} = \sum_{g_p \in C}
\ket{g_p} \bra{g_p}$ projects the plaquette state into the conjugacy class $C$ and is a gauge-invariant operator.

Therefore the magnetic term of the plaquette $p$ associates the energy $-\chi_j(C)/(2 \varg^2)$ to each conjugacy class $C$. The ground state of the magnetic Hamiltonian is defined by a configuration in which all the plaquettes are in the state associated with the identity operator, because the identity is the only group element which maximizes $\chi_j(C)$ to the value ${\rm dim}(j)$. All the other conjugacy classes, instead, define different kinds of localized magnetic vortices that are localized excitations over the ground states associated with a mass proportional to ${\rm dim}(j) - \chi_j(C)$.

In the case of finite groups, the Hamilonian $H_B$ defines a system with topological order (see, for example, \cite{fradkin} and references therein). In particular, the ground state for the gauge field defined by the Hamiltonian $H_B$ in the gauge-invariant sector and in the absence of matter corresponds to the ground state of a quantum double model \cite{Kitaev2003} associated to the same group. Also the excited states, defined by the presence of magnetic vortices, are the same in both models. The quantum double Hamiltonian, however, associates the same energy gap to all the species of magnetic vortices since it is constituted by the projector $\Pi_{1,p}$ over the identity only. The main difference in the magnetic sector of the two Hamiltonians is therefore constituted by the spectrum of the excitations. Furthermore the quantum double model implements dynamically the gauge invariance, which may be violated by the presence of gapped charged excitations. These excitations would play the role of static charges in the Kogut-Susskind Hamiltonian, where, instead, the matter particles are associated to charge excitations transforming under the group representation $j$.

Hence we recover the well known fact that the weak limit, $\varg \to 0$, of the Kogut-Susskind Hamiltonian for finite gauge groups is related to the corresponding quantum double model. The pure-gauge theory (without matter) is represented in this limit by a topological model in which the charge excitations have infinite mass, whereas the presence of matter in the gauge theory corresponds to a finite mass for the charge excitation sharing the same group representation.

Quantum double models are a particular example of string-net models \cite{Levin2005}. In particular the mapping from quantum double to the string-net models is obtained by interpreting the irreducible representations of the group as ``string types'' on the link of the string-net model \cite{Aguado2009}. Thus, these string types may be understood in terms of electric flux lines in the original lattice gauge theory. By exploiting the representation basis for the gauge fields, it is therefore possible to interpret the plaquette operators in $H_B$ in terms of the plaquette operators in the string-net model.

The plaquette operator \eqref{mag} transforms each link by effectively multiplying its state with the irreducible representation $j$ as described by Eq. \eqref{Uj}. This is a particular example of the plaquette operators appearing in the string net models which are, in general, associated to all the irreducible representations.
Therefore the Kogut-Sussking Hamiltonian reproduces one of the possible choices in the definition of the plaquette terms in the string-net model: the one in which only the representation $j$ appears. However, if $j$ is the fundamental representation of the group, the powers of the plaquette operator $W$ may be rewritten in terms of plaquette operators for all the other irreducible representations. This intuitively shows the equivalence of the magnetic part of the Hamiltonian and the string-net plaquette operators: they generate the same topological phase, as rigorously proved through the equivalence of the ground states with the quantum double model, evident in the group elements basis. In particular Levin and Wen \cite{Levin2005} argue that, for each finite gauge group, the string-net model in 2+1 dimensions provides an effective description of the scale invariant fixed point characterizing a deconfined phase in the associated lattice gauge theory, which may also be understood in terms of a condensation of the string-nets.

Beside finite groups it is also interesting to recall the form of the plaquette interactions in the continuous $U(1)$ theory (compact QED). In this case the group elements are simply phases. Thus one obtains the usual Abelian Kogut-Susskind magnetic energy,
\begin{equation}
H_B = -\frac{1}{\varg^2} \underset{p}{\sum} \cos \left( \phi_1 + \phi_2 - \phi_3 -\phi_4 \right)
\label{magab}
\end{equation}
Recall that the group elements (or more precisely, for compact groups, the group parameters) correspond to the vector potential. One can see that the argument of the cosine is a lattice, discrete curl, and thus expansion of the cosine for a continuum limit will result in a $\mathbf{B}^2$ term, corresponding to the usual Quantum-Electrodynamic magnetic energy term \cite{Kogut1979}.

\section{Summary}

We have described a method to construct lattice gauge theories out of basic atomic-like ingredients. This applies both for continuous, compact, Lie groups
(Kogut-Susskind Hamiltonian theories) as well as finite (discrete) groups. This may provide the background for both quantum simulations and new numerical approaches for such theories.

The matter degrees of freedom are represented by fermions, occupying the vertices of the lattice, and described by usual second-quantization Fock space fermionic operators - spinors in \emph{Group/Gauge space}. Other possible quantum numbers of the matter - such as spin and flavor - are independent of the gauge symmetry and thus were neglected, but they may be introduced in a straightforward way if required by the analysis of a specific model.

The gauge field degrees of freedom, on the other hand, are slightly different than in the ``conventional'' formulations. These are described by local Hilbert spaces on the links of the lattice, consisting of a set of possible modes with a total occupation number one. Thus a Fock space is not strictly required for these local Hilbert spaces. Nevertheless we described the link dynamics in terms of annihilation and creation operators, both for ease of notation and to provide a useful description for potential quantum simulations of the theories. Traditionally we refer to the gauge fields as bosons, but since their number operators do not exceed one, and only pairs of such operators are involved in the dynamics, their statistics does not play a role and thus they may be replaced by fermions as well.

These ``bosonic'' modes, or particles, correspond to the different \emph{representation states} of the gauge group. Hence, for finite groups, the local bosonic Hilbert spaces are finite, while for other groups the full Hilbert space is infinite, but it may be consistently truncated to a few representation states, as long as all the modes belonging to the participating representations are included. An analogous approach is adopted in \cite{Tagliacozzo2014} to truncate the tensors used to describe the gauge fields in a tensor network setting.

We have also reviewed the possibility to represent the group states in a different basis - the \emph{group element basis}, connected to the \emph{representation basis} by a unitary transformation which is a generalized Fourier transform, with the introduction of the generalized Wigner matrices. As the representation plays the role of an electric field, we deduce that group elements, as the conjugate quantum numbers, play the role of the vector potential. Indeed, the representation basis is more suitable for describing theories with matter, which is always in a given representation, and is related to the gauge invariance through the Gauss law, generated by the electric fields.

On the other hand, the group element basis is useful to analyze the content of the magnetic sector of the gauge theory and, in the case of discrete groups, allows the mapping of the weak-coupling limit of the theory into a quantum double model where the excitations of the plaquette terms correspond to magnetic vortices described by the conjugacy classes of the gauge group.

A similar mapping between these bases has been adopted in the tensor network description of gauge invariant states in \cite{Tagliacozzo2014}.

The lattice system has been described in terms of a square lattice, for various reasons: first, to relate this study to conventional high energy physics lattice gauge theories; second, to allow for simple generalization to any dimension (note that the dimension of the system was not addressed in the general framework); third, but not less important, for the sake of simplicity. However, generalizations to other lattices should be straightforward (one should note that bipartite lattices are required for the staggering of fermions, although other approaches for lattice fermions may also be utilized).

\section*{Acknowledgements.} The authors wish to thank M.C. Ba\~nuls, J.I. Cirac, M. Gurevich, S. K\"uhn, L. Tagliacozzo and T. Wahl for helpful discussions.
E.Z. acknowledges the support of the Alexander-von-Humboldt Foundation.

\section*{Appendix: Some examples} \label{sec:Example}

\subsection{A finite group example: $D_3$}
In order to summarize the full construction of the Kogut-Susskind Hamiltonian we present here the example provided by the smallest discrete non-Abelian group: the dihedral group $D_3$. The six elements of the group are generated by the rotation $\xi_{2\pi/3}$ of the angle $2\pi/3$ and the inversion symmetry $\sigma$ of the equilateral triangle. In particular it admits a fundamental representation $j=2$ in terms of $2 \times 2$ orthogonal matrices:
\begin{align}
 D^{(2)}\left( \xi_\alpha\right) &=\begin{pmatrix} \cos \alpha & \sin \alpha \\ -\sin \alpha & \cos\alpha \end{pmatrix} \,, \\
 D^{(2)}\left( \xi_\alpha\sigma\right) &=\begin{pmatrix} \cos \alpha & -\sin \alpha \\ -\sin \alpha & -\cos\alpha \end{pmatrix} \,,
\end{align}
with $\alpha=0,2\pi/3,4\pi/3$.
The other two irreducible representations are the trivial representation $D^I(g)=1$ and the parity representation $D^p(g)=\det(g)$.
Six states are therefore required to define the link Hilbert space and from the representation matrices is straightforward to obtain the left and right transformation operators $\Theta^L_g$ and $\Theta^R_g$ in Eqs. (\ref{Ltransf},\ref{Rtransf}).

The particle fields transform under the $j=2$ representation and can be described through the spinor $\vec{\psi}=(\psi_\uparrow,\psi_\downarrow)$.
The corresponding charge operators in Eq. \eqref{theta} read:
\begin{multline}
\Theta^{Q,2}_{\xi_\alpha} = e^{\alpha\left(\psi_\Up^\dag \psi_\Dn - \psi^\dag_\Dn\psi_\Up \right)} = \\ =
1 - \left(1-\cos{\alpha} \right)\left(n_\Up + n_\Dn \right) + \sin\alpha\left(\psi_\Up^\dag \psi_\Dn - \psi^\dag_\Dn\psi_\Up \right)+ \\ + 2\left(1-\cos\alpha \right)n_\Up n_\Dn
\end{multline}
for the rotations having determinant equal to 1, and
\begin{multline}
\Theta^{Q,2}_{\xi_\alpha\sigma} =  \left[ 1 + \left(\cos\alpha -1\right)n_\Up - \left(\cos \alpha+1 \right)n_\Dn + \right. \\
\left.+ \sin\alpha\left(\psi_\Up^\dag \psi_\Dn + \psi^\dag_\Dn\psi_\Up \right)\right] (-1)^N
\end{multline}
where we adopted the notation $n_i=\psi^\dag_i\psi_i$. In particular $\Theta^{Q,2}_{\sigma}=(1-2n_\Dn)(-1)^N$.

From the previous operators it is easy to define the Gauss law on each vertex of the lattice. The operators $U^{(2)}_{mm'}$ in \eqref{CGS} can be evaluated, instead, from the Clebsch-Gordan coefficients (calculated using the technique in \cite{clebsch}):
\begin{align}
\bracket{I,2m}{2n}&=\delta_{mn} \,,\\
\bracket{p,2m}{2n}&=\epsilon_{mn} \,,\\
\bracket{2n,2m}{I}&=\delta_{mn}/\sqrt{2} \,,\\
\bracket{2n,2m}{p}&=\epsilon_{nm}/\sqrt{2} \,,\\
\bracket{2n,2m}{2l}&=\left(\delta_{l\Up}\sigma^z_{nm}-\delta_{l\Dn}\sigma^x_{nm}\right)/ \sqrt{2} \,,
\end{align}
where $\sigma^x$ and $\sigma^z$ are Pauli matrices for the indices $n,m$.

Finally the spectrum of the plaquette term for the group $D_3$ is composed by three conjugacy classes corresponding to the identity, the rotations of $2\pi/3$ and $4\pi/3$ and the inversions. A detailed analysis of the quantum double model for this group may be found, for example, in \cite{Brennen2009}.

\subsection{A compact Lie group example: $SU(2)$} \label{sec:SU2EX}

As an example of a compact Lie group, we examine the explicit construction of an $SU(2)$ lattice gauge theory.

First, let us consider the structure of the Hilbert space of the gauge field in that case. For this, it will be instructive to review the rigid rotator analogy made by Kogut and Susskind
in their first paper about Hamiltonian lattice gauge theory \cite{KogutSusskind}.
As discussed in section II, the $D$ matrices are simply generalizations of the Wigner matrices. These are, up to a normalization, eigenfunctions of the rigid rotator problem. Recall that a rigid rotator configuration space is three dimensional, and may be described either by three coordinates, like the Euler angles $\alpha, \beta, \gamma$ which parametrize either the configuration space or simply the $SU(2)$ group elements, or by three integers $j,m,n$, where $m$ is the eigenvalue of the $z$ component of the angular momentum in the space frame, $J^s_z$, $n$ is the eigenvalue of the $z$ component of the angular momentum in the body frame, $J^b_z$, and $j$ is the total angular momentum quantum number, shared by both frames,
$\mathbf{J}^2 = \left(\mathbf{J}^s\right)^2 = \left(\mathbf{J}^b\right)^2$. The angular momentum operators satisfy the algebra \cite{Landau1981,KogutSusskind}
$\left[J^{\alpha}_i,J^{\beta}_j\right] = i \delta^{\alpha \beta} (-1)^{\delta _{\alpha b}} \epsilon_{ijk}J^{\alpha}_k$
with $\alpha, \beta = s,b$ and $i,j,k=1,2,3$ or $x,y,z$.
The $D$ matrices are merely the wavefunctions \cite{Landau1981,Rose1995}:
\begin{equation}
\left\langle \alpha, \beta, \gamma | j m n \right\rangle = \sqrt{\frac{2j+1}{8 \pi^2}} D^{j} _{mn} \left(\alpha, \beta, \gamma\right).
\end{equation}
This change of basis may be seen as a generalized, non-Abelian Fourier transform, mapping between the \emph{Representation States} $\left| j m n \right\rangle$ and
\emph{Group Element States}  $\left| \alpha, \beta, \gamma \right\rangle$.

One may generalize this relation to any group $G$ by replacing $\alpha, \beta, \gamma$ by the general $g \in G$, resulting in equation (\ref{Fourier}).

Finally, let us consider the analogs of the space and body angular momenta. Let us go back to the rigid rotator problem. Denote by $U^{j}$ a rotation matrix (group element) expressed
 in the representation $j$. This satisfies the following commutation relations with the angular momentum operators: $\left[J^b_i,U_{mn}^{j}\right]=(T^{j}_{i})_{mk}U_{kn}^{j}$
 and $\left[J^s_i,U_{mn}^{j}\right]=U_{mk}^{j}(T^{j}_{i})_{kn}$ with $T_i^j$ being the $j$th representation of the $i$ component of angular momentum. That means that the body-frame angular momenta generate left transformations on group elements, and the ones of the space-frame - right ones. Recall that the body and space algebras had opposite signs for the structure constants, and if we wish to have the same sign in both of the algebras, we can simply define $R_i \equiv J^s_i$ and $L_i \equiv -J^b_i$. Then, one obtains the $SU(2)$ algebra
 described by equation (\ref{algebra}) with $f_{abc}=\epsilon_{abc}$.

After understanding the Hilbert space structure, and following equation (\ref{NAcharge}), we can obtain an explicit form for the charge operator: for $j=1/2$, the fundamental representation of $SU(2)$, for example,  we have $\mathbf{T}^{1/2} = \vec{\sigma}/2$, and thus we obtain the charges
\begin{equation}
\mathbf{Q}^{1/2}_{SU(2)} = \frac{1}{2}\psi^{\dagger}_a \vec{\sigma}_{ab} \psi_b
\end{equation}
as in \cite{KogutSusskind}. It is easy to check that for both a full and an empty vertex, the charge is zero (a singlet state), in accordance with the fact that there is only one trivial representation.

The generators, or electric field operators, are formed by plugging the representation matrices of $SU(2)$ into the general prescriptions given in equations (\ref{Lgen}) and (\ref{Rgen}).
The matrix elements of $U^{j}$ may be easily calculated using the Clebsch-Gordan coefficients. In equation (\ref{CGS}), one simply has to substitute $\text{dim}\left(J\right)=2J+1,\text{dim}\left(K\right)=2K+1$. It is particularly simple for $j=1/2$, where $K=J \pm 1/2$, and then it is straightforward to obtain:
 \begin{align}
  \left\langle  J, M, 1/2, m \,|\, J+1/2, N \right\rangle &= \sqrt{\frac{J+2mM+1}{2J+1}} \delta_{N,M+m} \,,\\
  \left\langle   J, M, 1/2, m \,|\, J-1/2, N \right\rangle &= -2m\sqrt{\frac{J-2mM}{2J+1}} \delta_{N,M+m}\,.
 \end{align}
If one wishes to truncate the theory for some $j \leq J_{max}$, only the first $\left\{\left|jmn\right\rangle\right\}_{j=0}^{J_{max}}$ states ($\sum_{j=0}^{J_{max}}\left(2j+1\right)^2 = \left(J_{max}+1\right)\left(2J_{max}+1\right)\left(4J_{max}+3\right)/3$ states in total) are used,
 and the corresponding series (generators and $U$ matrix elements) should be truncated accordingly. Then, the trace of the $U^{j}$ matrix is given by
 \begin{equation}
\text{Tr}\left(U^{j\dagger}U^j\right) = \text{tr}\left(U^jU^{j\dagger}\right) = 2j+1 - f_j\left(J_{max}\right)P_{J_{max}}
\label{trace}
\end{equation}
where $P_{J_{max}}$ projects to the subspace of $J_{\max}$. Note that the trace is an \emph{operator}, but yet it is gauge invariant. For $j=1/2$, one obtains $f_{1/2}\left(J\right) = (2J+2)/(2J+1)$.

 As an example, if the truncation includes only the two lowest representations, $j=0,1/2$, only five states $\left|jmn\right\rangle$ and their corresponding creation/annihilation operators are required. In that case, the generators take the form
\begin{equation}
\mathbf{L} = - \frac{1}{2} a^{\dagger 1/2}_{ml} \vec{\sigma}_{nm} a^{1/2}_{nl}
\end{equation}
and
 \begin{equation}
\mathbf{R} =   \frac{1}{2} a^{\dagger 1/2}_{ml} \vec{\sigma}_{mn} a^{1/2}_{nl}
\end{equation}
and the matrix of operators $U^{1/2}$ is given by
\begin{equation}
U^{1/2}=\frac{1}{\sqrt{2}}\begin{pmatrix}
a_{\Up,\Up}^{1/2 \dagger}a^{0}_{00}+a_{00}^{0\dagger}a^{1/2}_{\Dn,\Dn} & a_{\Up,\Dn}^{1/2\dagger}a^{0}_{00}-a_{00}^{0\dagger}a^{1/2}_{\Dn,\Up}\\
a_{\Dn,\Up}^{1/2\dagger}a^{0}_{00}-a_{00}^{0\dagger}a^{1/2}_{\Up,\Dn} & a_{00}^{0\dagger}a^{1/2}_{\Up,\Up}+a_{\Dn,\Dn}^{1/2\dagger}a^{0}_{00}
\end{pmatrix}\,,
\end{equation}
where we used the shorthand notation $\Up=1/2$ and $\Dn=-1/2$.

This matrix acts within a five-dimensional Hilbert space, as in the $SU(2)$ link model \cite{Horn1981,Orland1990,Wiese1997,Brower2004} (see also \cite{Tagliacozzo2014}), and thus this particular truncation is similar. According to the link model, such a truncation is possible using $so(5)$ as an embedding algebra for $SU(2)$. However, more generally, the method presented in this paper allows for other truncations as well (larger values of $J_{max}$) - these are representation-based truncations, unlike (in general) in the link model. The current model allows to restore the complete Kogut-Susskind theory by enlarging the truncation, rather than obtaining the continuum limit using an extra compact dimension as in the link model.

Another difference from the link model, and also from the prepotential approach \cite{MathurSU2,Mathur2007}, is that the gauge degrees of freedom do not have to be separated into left and right spaces, but the link is rather taken as a whole piece. That should increase the feasibility of quantum simulations.

Furthermore, in the prepotential approach, the construction of the $U$ operator involves both terms which violate the number conservation of the particles on the links, and square roots of number operators in its denominator. This sets a challenge in quantum simulating the model (see the quantum simulation proposal \cite{NA}, for example). In our proposal, instead, we use only a single particle to represent the gauge field in a link and the denominator operators are absent. This is due to the different way of representing the gauge fields degrees of freedom, namely the mapping to a Schwinger algebra in the prepotential formalism and the use of single-particle modes here.


\begin{thebibliography}{50}
\bibitem{footnote} Keeping in mind that for general groups the Clebsch-Gordan coefficients are unnecessarily real, we use the rather general form
\unexpanded{$D^{j_1}_{m_1 m_1'} D^{j_2}_{m_2 m_2'} = \sum_J \bracket{j_1 m_1 j_2 m_2}{J M} \bracket{J M'}{j_1 m_1' j_2 m_2'} D^{J}_{MM'}$}
\end{thebibliography}
\end{document}